\documentclass[prl,aps,amsfonts,amsmath,amssymb,nofootinbib,twocolumn,superscriptaddress]{revtex4}

\usepackage{graphicx}
\usepackage{bm}
\usepackage{hyperref}
\usepackage{IEEEtrantools}
\usepackage{xcolor}

\usepackage{natbib}
\usepackage{float}

\restylefloat{table}

\begin{document}

	\title{Low-energy electrodynamics of infinite-layer nickelates: evidence for \textit{d}--wave superconductivity in the dirty limit}
	
	\affiliation{Ames National Laboratory, Ames, IA 50011 USA}
	
	\affiliation{Stanford Institute for Materials and Energy Sciences, SLAC National Accelerator Laboratory, Menlo Park, CA 94025, USA}
	
	\affiliation{Department of Physics, Stanford University, Stanford, CA 94305, USA}

	\affiliation{Department of Applied Physics, Stanford University, Stanford, CA 94305, USA}
	
	\affiliation{Department of Physics, University of Alabama at Birmingham, Birmingham, AL 35294-1170, USA.}

     \affiliation{Department of Physics and Astronomy, Iowa State University, Ames, Iowa 50011, USA.}

	\author{Bing Cheng}
	\email{bcheng2@ameslab.gov}
	\affiliation{Ames National Laboratory, Ames, IA 50011 USA}

	\author{Di Cheng}
	\affiliation{Ames National Laboratory, Ames, IA 50011 USA}

	\author{Kyuho Lee}
	\affiliation{Stanford Institute for Materials and Energy Sciences, SLAC National Accelerator Laboratory, Menlo Park, CA 94025, USA}
	\affiliation{Department of Physics, Stanford University, Stanford, CA 94305, USA}

	\author{Liang Luo}
	\affiliation{Ames National Laboratory, Ames, IA 50011 USA}
	
	\author{Zhuoyu Chen}
	\affiliation{Stanford Institute for Materials and Energy Sciences, SLAC National Accelerator Laboratory, Menlo Park, CA 94025, USA}
	\affiliation{Department of Applied Physics, Stanford University, Stanford, CA 94305, USA}
	
	\author{Yonghun Lee}
	\affiliation{Stanford Institute for Materials and Energy Sciences, SLAC National Accelerator Laboratory, Menlo Park, CA 94025, USA}
	\affiliation{Department of Applied Physics, Stanford University, Stanford, CA 94305, USA}

	\author{Bai Yang Wang}
	\affiliation{Stanford Institute for Materials and Energy Sciences, SLAC National Accelerator Laboratory, Menlo Park, CA 94025, USA}
	\affiliation{Department of Physics, Stanford University, Stanford, CA 94305, USA}

	\author{Martin Mootz}
	\affiliation{Ames National Laboratory, Ames, IA 50011 USA}
	\affiliation{Department of Physics and Astronomy, Iowa State University, Ames, Iowa 50011, USA.}

	\author{Ilias E. Perakis}
	\affiliation{Department of Physics, University of Alabama at Birmingham, Birmingham, AL 35294-1170, USA.}
	
	\author{Zhi-Xun Shen}
	\affiliation{Stanford Institute for Materials and Energy Sciences, SLAC National Accelerator Laboratory, Menlo Park, CA 94025, USA}
	\affiliation{Department of Physics, Stanford University, Stanford, CA 94305, USA}
	\affiliation{Department of Applied Physics, Stanford University, Stanford, CA 94305, USA}
	
	\author{Harold Y. Hwang}
	\affiliation{Stanford Institute for Materials and Energy Sciences, SLAC National Accelerator Laboratory, Menlo Park, CA 94025, USA}
	\affiliation{Department of Applied Physics, Stanford University, Stanford, CA 94305, USA}

	\author{Jigang Wang}\email{jgwang@ameslab.gov}
	\affiliation{Ames National Laboratory, Ames, IA 50011 USA}
	\affiliation{Department of Physics and Astronomy, Iowa State University, Ames, Iowa 50011, USA.}

\date{\today}

\begin{abstract}

\textbf{The discovery of superconductivity in infinite-layer nickelates establishes a new category of unconventional superconductors that share structural and electronic similarities with cuprates. Despite exciting advances, such as the establishment of a cuprate-like phase diagram and the observation of charge order and short-range antiferromagnetic fluctuation, the key issues of superconducting pairing symmetry, gap amplitude, and superconducting fluctuation remain elusive. In this work, we utilize static and ultrafast terahertz spectroscopy to address these outstanding problems. We demonstrate that the equilibrium terahertz conductivity and nonequilibrium terahertz responses of an optimally Sr-doped nickelate film ($T_c$ = 17 K) are in line with the electrodynamics of $d$-wave superconductivity in the dirty limit. The gap-to-$T_c$ ratio 2$\Delta/k_\mathrm{B}T_\mathrm{c}$ is extracted to be 3.4, indicating the superconductivity falls in the weak-coupling regime. In addition, we observed significant superconducting fluctuation near $T_\mathrm{c}$, while it does not extend into the deep normal state as optimally hole-doped cuprates. Our result highlights a new $d$-wave system which closely resembles the electron-doped cuprates, expanding the family of unconventional superconductivity in oxides.} 

\end{abstract}

\maketitle

\setlength{\parskip}{0.1em}

The complexity arising from the strong electronic correlation in high-temperature superconductors, such as the cuprates, goes far beyond the celebrated Landau Fermi liquid and Bardeen-Cooper-Schrieffer (BCS) theories\cite{RMP_review_wen_2006}. Despite several decades of research effort devoted, the fundamental problem of how electrons bind and condense remains elusive\cite{cuprates_review_nature_2015}. The recent discovery of infinite-layer nickelate superconductors, which share structural and electronic similarities with cuprates, provides a unique opportunity to decipher this long-standing mystery\cite{nickelate_nature_2019}. Thus far, experimental progress on nickelate electronic and magnetic structures has been primarily achieved by the measurements from X-ray based techniques, such as X-ray photoemission spectroscopy and resonant inelastic X-ray scattering, which reveal a Mott-Hubbard scenario for electronic structure\cite{RXIS_nickelate_NM_2020,CHEN20221806}, as well as short-range antiferromagnetic fluctuations in magnetic structure of both parent and doped nickelate films\cite{RXIS_magntic_2021}.

Despite these progresses, to date, the key open questions regarding superconducting pairing symmetry and the gap amplitude in nickelates remain under intense debate and have not yet been unambiguously determined\cite{nickelate_dwaveprb_theory_2020,nickelate_dwaveprl_theory_2020,nickelate_pen_depth_Huang_2022,nickelate_pen_depth_Zhang_2022}. In general, crucial information about the structure and symmetry of superconducting gap could be from the photoemission measurements, which are able to reveal the gap anisotropy in the momentum space\cite{dwave_arpes_1993}, and the scanning SQUID measurements, which are sensitive to the phase of superconducting order parameter \cite{dwave_nature_1995}. However, the \emph{ex situ} chemical reduction process at sample synthesis alters the atomically flat surface of the grown nickelate films, making it difficult for these surface-sensitive techniques to probe the superconducting gap structure and symmetry in nickelate films. To propel the field forward, a bulk-sensitive spectroscopic measurement of nickelate films, which is able to complete these critical missing pieces, is highly demanded.

\begin{figure*}[t]

\includegraphics[clip,width=4.6in]{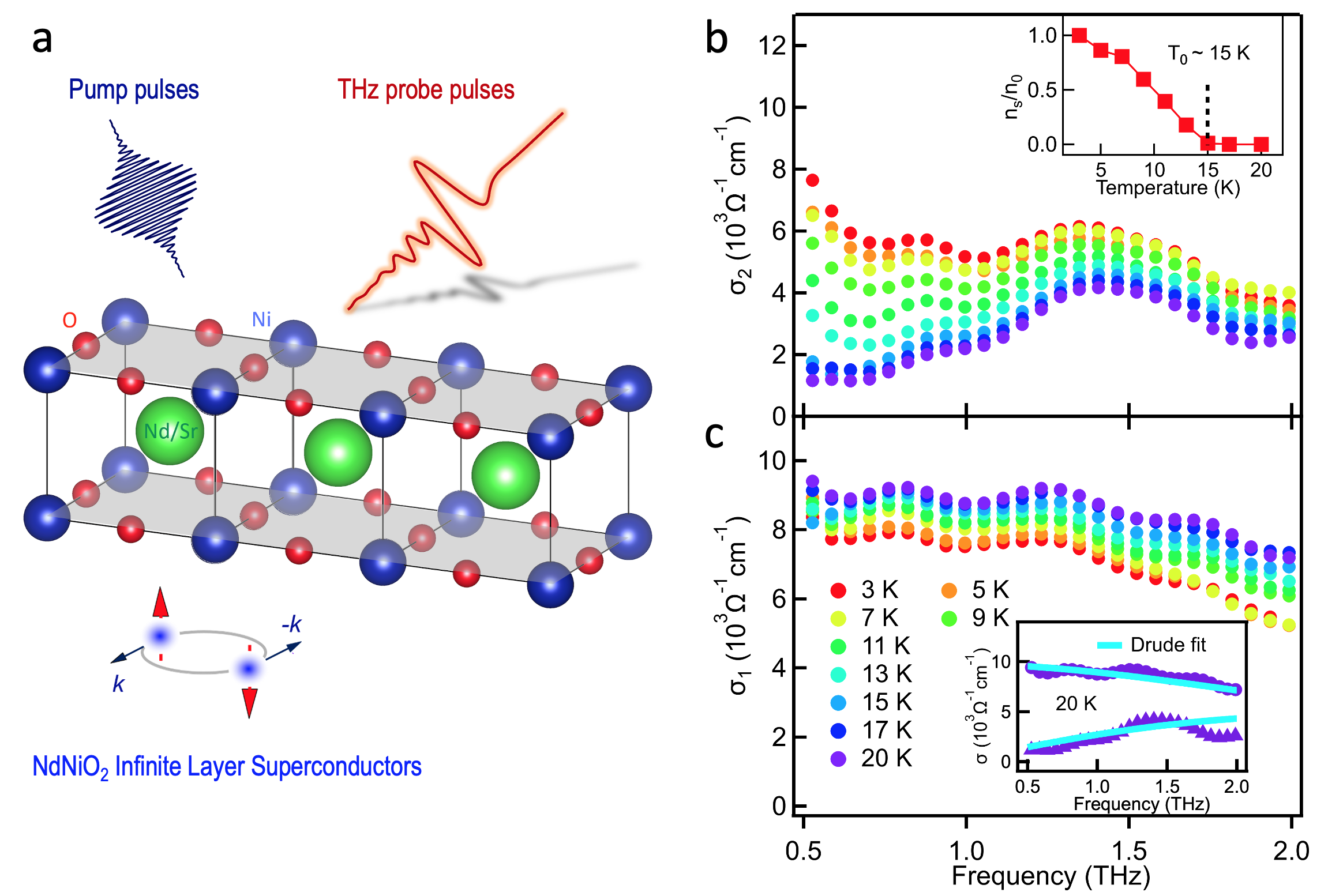}
\caption{ {\bf Equilibrium THz conductivity of nickelate superconducting film Nd$_{0.85}$Sr$_{0.15}$NiO$_2$.}  {\bf (a)} Schematic illustration of the lattice structure of Nd$_{0.85}$Sr$_{0.15}$NiO$_2$ and our probe schemes of static and ultrafast THz spectroscopy. {\bf (b)} The imaginary part of THz conductivity $\sigma_{2}(\omega)$ at a few temperatures. The inset shows the normalized superfluid density as a function of temperature extracted by the two-fluid model analysis.  {\bf (c)} The real part of THz conductivity $\sigma_{1}(\omega)$ at a few temperatures. The inset shows a Drude model fit to the normal-state THz conductivity.}

\label{Fig1}
\end{figure*}

Terahertz (THz) electrodynamic measurement is an extremely \textit{bulk-sensitive} probe that can overcome the limitations imposed by the nickelate film synthesis process and reveal the genuine dynamical superconducting response\cite{swave_cheng_2016,THz_sc1,THz_sc2}. In this work, we utilize time-domain THz spectroscopy and optical-pump, THz probe to study the superconducting gap structure and superconducting fluctuation in an optimally Sr-doped nickelate film Nd$_{0.85}$Sr$_{0.15}$NiO$_2$ ($T_\mathrm{c}$ = 17 K)\cite{nickelate_LAST_2022}. These nickelate films exhibit excellent crystallinity and minimal defects, as confirmed by the X-ray diffraction, scanning transmission electron microscopy, and dc resistivity measurements\cite{nickelate_LAST_2022}. Fig. 1a illustrates the lattice structure of nickelate film as well as our probe schemes of static and ultrafast THz spectroscopy. We demonstrate that the complex terahertz conductivity below $T_c$, both in and out of equilibrium, are in line with the electrodynamics of dirty-limit $d$-wave superconductivity. We further extract the superconducting gap amplitude and  demonstrate significant superconducting fluctuation near $T_c$. Our findings suggest that the superconductivity in the nickelate film is alike the $d$-wave superconductivity in electron-doped cuprates.

\begin{figure*}[t]

\includegraphics[clip,width=5.5in]{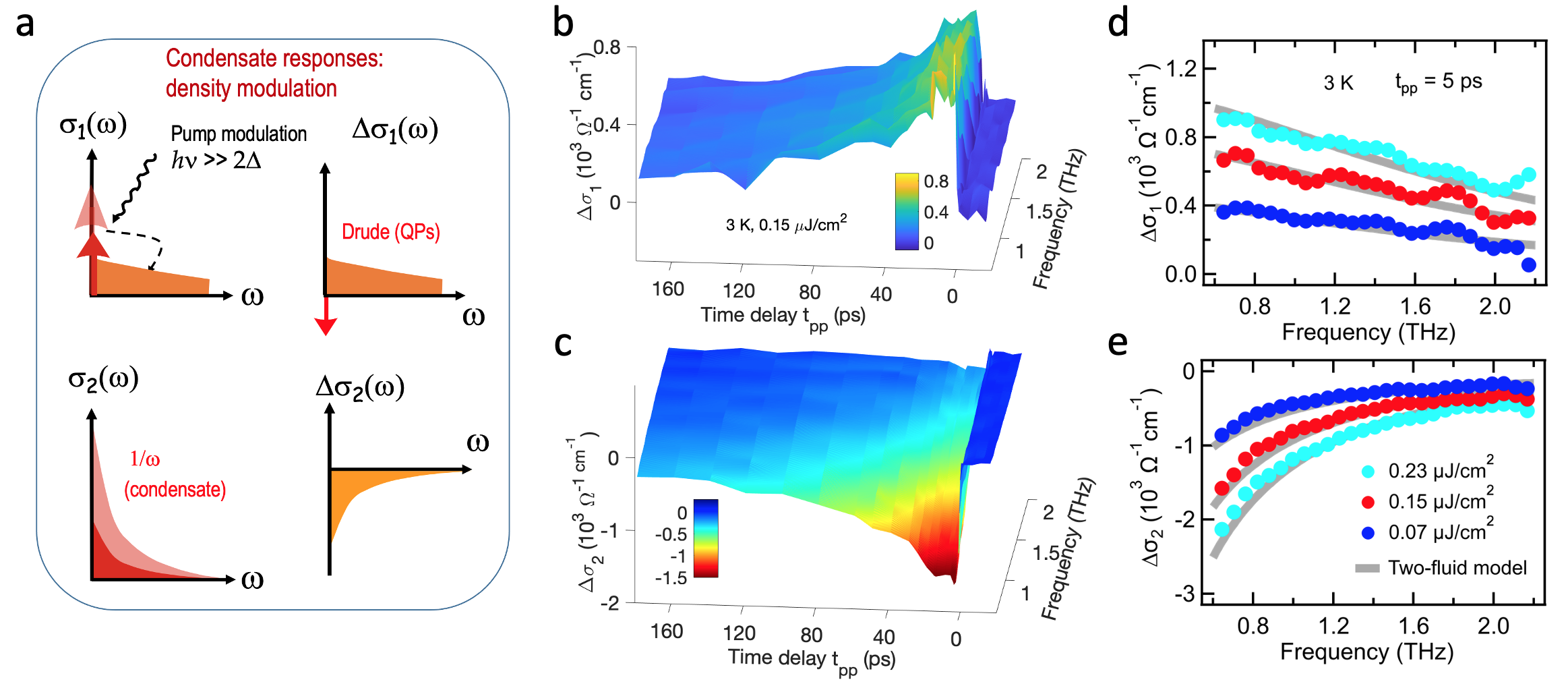}
\caption{ {\bf Ultrafast THz conductivity $\Delta\sigma(t_\mathrm{pp}, \omega)$ at 3 K.}  {\bf (a)} Schematic illustration of how the superconducting condensate is modulated by optical excitation. A weak optical pump will break a small fraction of Cooper pairs, weakening the $1/\omega$ divergence of $\sigma_2(\omega)$. As a result, $\sigma_1(\omega)$ is enhanced by the extra Drude transport of photoexcited quasiparticles. $\Delta\sigma_{2}(\omega)$ will show a negative $1/\omega$ divergence, $\Delta\sigma_{1}(\omega)$ will show a positive Drude-like peak in $d$-wave superconductor. {\bf (b)} The three-dimensional plot of the real part of ultrafast THz conductivity $\Delta\sigma_{1}(\omega,t_\mathrm{pp})$ and {\bf (c)} imaginary part of ultrafast THz conductivity $\Delta\sigma_{2}(\omega,t_\mathrm{pp})$ at 3 K. The pump fluence is 0.15 $\mathrm{\mu J}$/cm$^2$. {\bf (d)} The real part of ultrafast THz conductivity $\Delta\sigma_{1}(\omega)$ and {\bf (e)} imaginary part of ultrafast THz conductivity $\Delta\sigma_{2}(\omega)$ as a function of frequency at $t_\mathrm{pp} = 5$ ps under three pump fluences 0.07, 0.15, 0.23 $\mathrm{\mu J}$/cm$^2$. The gray curves are two-fluid model simulations to $\Delta\sigma_{1}$ and $\Delta\sigma_{2}$.}

\label{Fig2}
\end{figure*}

\noindent{\textbf{Equilibrium THz electrodynamics across superconducting transition}}

The imaginary and real parts of THz conductivity, $\sigma_2(\omega)$ and $\sigma_1(\omega)$, are shown in Fig. 1b and 1c at different temperatures respectively. In the normal state, $\sigma_1$ shows a typical Drude-like peak and $\sigma_2$ is a simple increasing function of frequency, consistent with the metallic behavior revealed by dc transports\cite{nickelate_LAST_2022}. The normal-state Drude scattering rate $\gamma_\mathrm{D}$ at 20 K, which represents the impurity scattering strength in the nickelate film, is extracted to be $\sim$ 3.3 THz [inset, Fig. 1c]. This value is in the same order of the impurty scattering (2 $\sim$ 4 THz) in La$_{2-x}$Sr$_x$CuO$_4$, NbSn$_3$, and Ba$_2$(Fe$_{1-x}$Co$_x$)As$_2$ films\cite{LSCO_THzfilm_2019,THz_sc1,THz_sc2}, further attesting to the high quality of our films. Across $T_\mathrm{c}$ to the superconducting state, $\sigma_1$ preserves the Drude-like line shape but gradually loses spectral weight, corresponding to the opening of superconducting gap. The missing spectral weight of $\sigma_1$ condenses into a delta function at zero frequency and consequently, the reactive superconducting condensate gives rise to a $1/\omega$ divergence on $\sigma_2$ as witnessed in the 3-15\,~K measurements in Fig. 1b. We emphasize two striking observations in conductivity spectra towards the $T$ $\rightarrow$ 0 limit: (1) absence of $s$-wave like excitation gap structure in $\sigma_1$, and (2) a significant residual Drude peak in $\sigma_1$.

In conventional dirty-limit \textit{s}-wave superconductors, the formation of a full superconducting gap on the Fermi surface forces $\sigma_1(\omega)$ to vanish below 2$\Delta$, and to rise steeply above 2$\Delta$ in a fashion that depends non-trivially on the BCS coherence factor\cite{swave_cheng_2016,THz_sc1}. Our observations (1) and (2) in the nickelate film evidently contradict with such a frequency dependence in $s$-wave superconductors. In contrast, in $d$-wave superconductors like cuprates, the $d$-wave gap symmetry enforces four gap nodes on the superconducting gap structure, leading to a gapless quasiparticle excitation spectrum. Without considering the pair-breaking effect from impurities, $\sigma_1(\omega)$ is approximately a linear increasing function of frequency below 2$\Delta$ at zero temperature\cite{dwave_linearcond_1994}. Importantly, the effect of impurity scattering, even if the strength is relatively moderate, could remarkably smear the $d$-wave gap nodes and create a significant number of nodal quasiparticles that eventually manifests themselves as a Drude peak at low frequency. In cuprates, no matter in high-quality single crystals or thin films, $\sigma_1(\omega)$ in superconducting state commonly shows a large uncondensed Drude component below 2$\Delta$ in the $T$ $\rightarrow$ 0 limit\cite{LSCO_infraredcrystal_2005,LSCO_THzfilm_2019,LCCO_THzfilm_2021}. Our observed low--energy conductivity spectra in nickelate film are very similar to the electrodynamics well-established in cuprates. Despite the exact origin of such significant uncondensed quasiparticles in cuprates is still under debate, the absence of any superconducting gap structures, and the presence of a significant uncondensed Drude component in the measured conductivity, provide clear experimental evidence to support a potential $d$-wave scenario in the nickelate film.

\noindent{\textbf{Ultrafast THz conductivity of superconducting condensate}}

Static THz measurement reveals mixed responses from superconducting condensate and uncondensed quasiparticles. To further investigate the $d$-wave scenario, we employed  optical-pump, THz probe, which exclusively probes the superconducting condensate and Cooper pair breaking response, to study the superconducting gap structure in the nickelate film. A weak 1.5 eV excitation of nickelate film will break a tiny fraction of Cooper pairs. In a $d$-wave system, this non-thermal depletion of superconducting condensate will enforce a \textit{negative} $1/\omega$ divergence on $\Delta\sigma_2(\omega)$ and a \textit{positive} Drude-like peak at low frequency on $\Delta\sigma_1(\omega)$, as illustrated by the schematics in Fig. 2a.

\begin{figure*}[t]

\includegraphics[clip,width=5.2in]{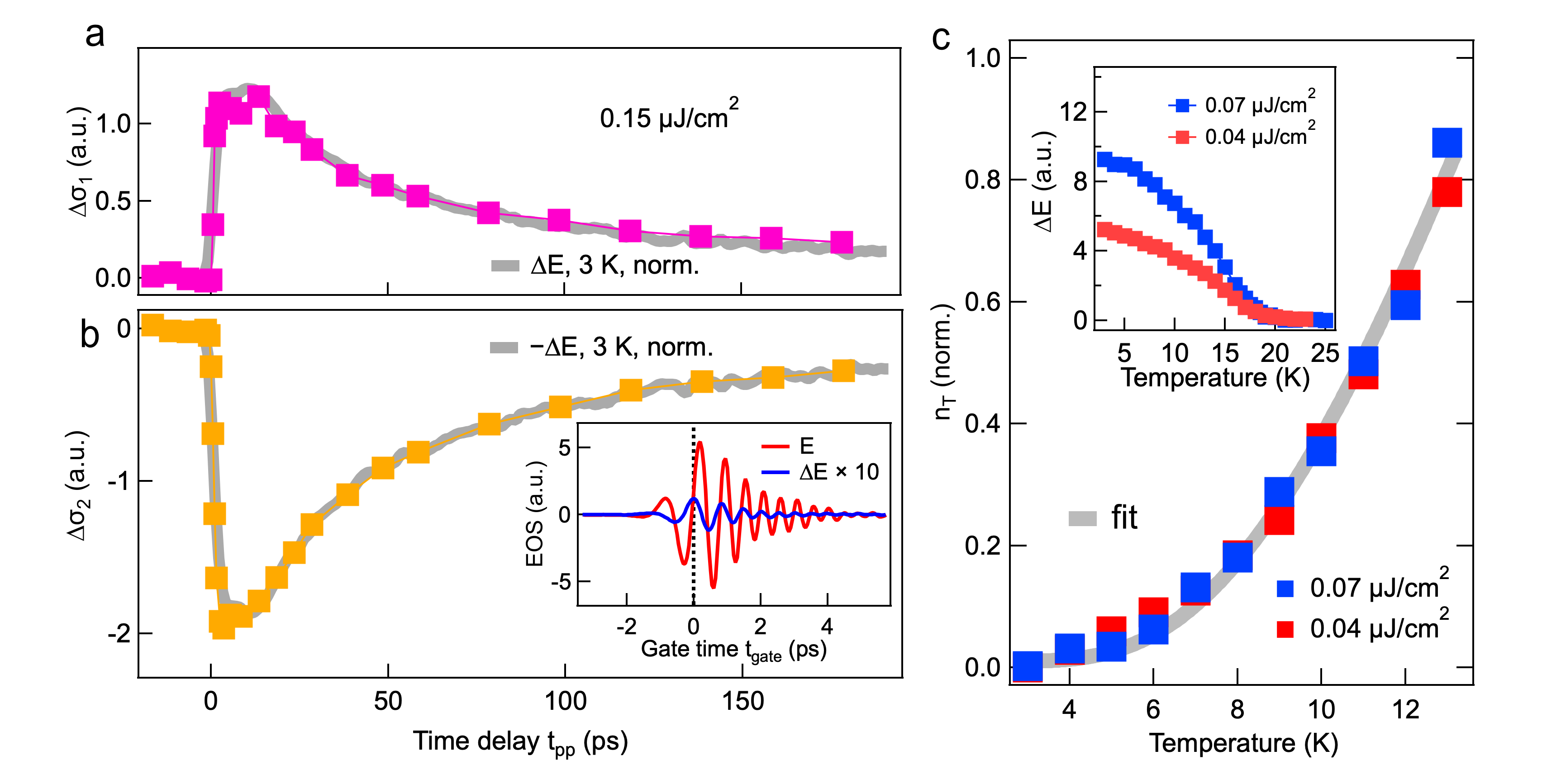}
\caption{ {\bf Ultrafast superconducting kinetics to extract superconducting gap 2$\Delta$.}  {\bf (a)} The real part of ultrafast THz conductivity $\Delta\sigma_1$ and {\bf (b)} The imaginary part of ultrafast THz conductivity $\Delta\sigma_2$ as a function of delay time $t_\mathrm{pp}$. The pump fluence is 0.15 $\mathrm{\mu J}$/cm$^2$. Here $\Delta\sigma_1(t_\mathrm{pp})$ and $\Delta\sigma_2(t_\mathrm{pp})$ are integrated over a 0.4 THz interval (centered at 0.95 THz) of the data in Fig. 2b and 2c respectively. The delay-time trace of pump-induced THz field change $\Delta E$ is also plotted along with $\Delta\sigma_1(t_\mathrm{pp})$ and $\Delta\sigma_2(t_\mathrm{pp})$. The inset shows the time trace of THz field in static measurement and the time trace of pump-induced THz field change in ultrafast measurement at 3 K. The vertical dash line is the position where we did $\Delta E$ scan vs time delay $t_\mathrm{pp}$. {\bf (c)} Normalized thermal quasiparticle density $n_T$ as a function of temperature measured under two weak pump fluences 0.04, 0.07 $\mathrm{\mu J}$/cm$^2$. The inset shows the temperature dependent $\Delta E$ measured at $t_\mathrm{pp}$ = 5 ps.}

\label{Fig3}
\end{figure*}

We present complex nonequilibrium THz conductivity $\Delta\sigma_1$ and $\Delta\sigma_2$ as functions of frequency and pump-probe time delay $t_\mathrm{pp}$ in Fig. 2b and 2c respectively. Inspecting from the $t_\mathrm{pp}$ axis, we observe the pump excitation causes an increase in $\Delta\sigma_1(t_\mathrm{pp})$ and a reduction in $\Delta\sigma_2(t_\mathrm{pp})$, both followed by a similar, slow recovery in a timescale of $\sim$ 100 ps. To highlight the frequency response, we plot $\Delta\sigma_1$ and $\Delta\sigma_2$ as a function of frequency at $t_\mathrm{pp}$ = 5 ps with high statistics in Fig. 2d and 2e respectively. One can see $\Delta\sigma_1(\omega)$ shows a well-defined \textit{positive} Drude-like peak, and $\Delta\sigma_2(\omega)$ displays a \textit{negative} $1/\omega$ divergence under all pump fluences. A two-fluid model $\Delta\sigma(\omega)$$=$$S_\mathrm{qp}\tau/(1-i\omega\tau)$$-$$S_\mathrm{sc}$$(\pi\delta(\omega)+i/\omega)$ is able to well reproduce $\Delta\sigma_1(\omega)$ and $\Delta\sigma_2(\omega)$. The representative simulations are plotted together as the gray curves in Fig. 2d and 2e. The analysis shows the pump depleted spectral weight of superconducting condensate $S_\mathrm{sc}$ has been fully transferred into the quasiparticle excitations $S_\mathrm{qp}$. Most intriguingly, $\Delta\sigma_1(\omega)$ does not inherit any $s$-wave like gap structures below 2.2 THz either, which serve as an additional evidence to exclude the possibility of s-wave gap symmetry\cite{THz_sc1,THz_sc3}. In the meanwhile, all our observations could be naturally understood in the context of the $d$-wave scenario. Similar Drude-like transport on $\Delta\sigma_1(\omega)$ below 2$\Delta$ has been commonly observed in the nonequilibrium superconducting states of cuprates\cite{pumped_cuprates_2001,pumped_cuprates_2005}.

\begin{figure*}[t]

\includegraphics[clip,width=5.5in]{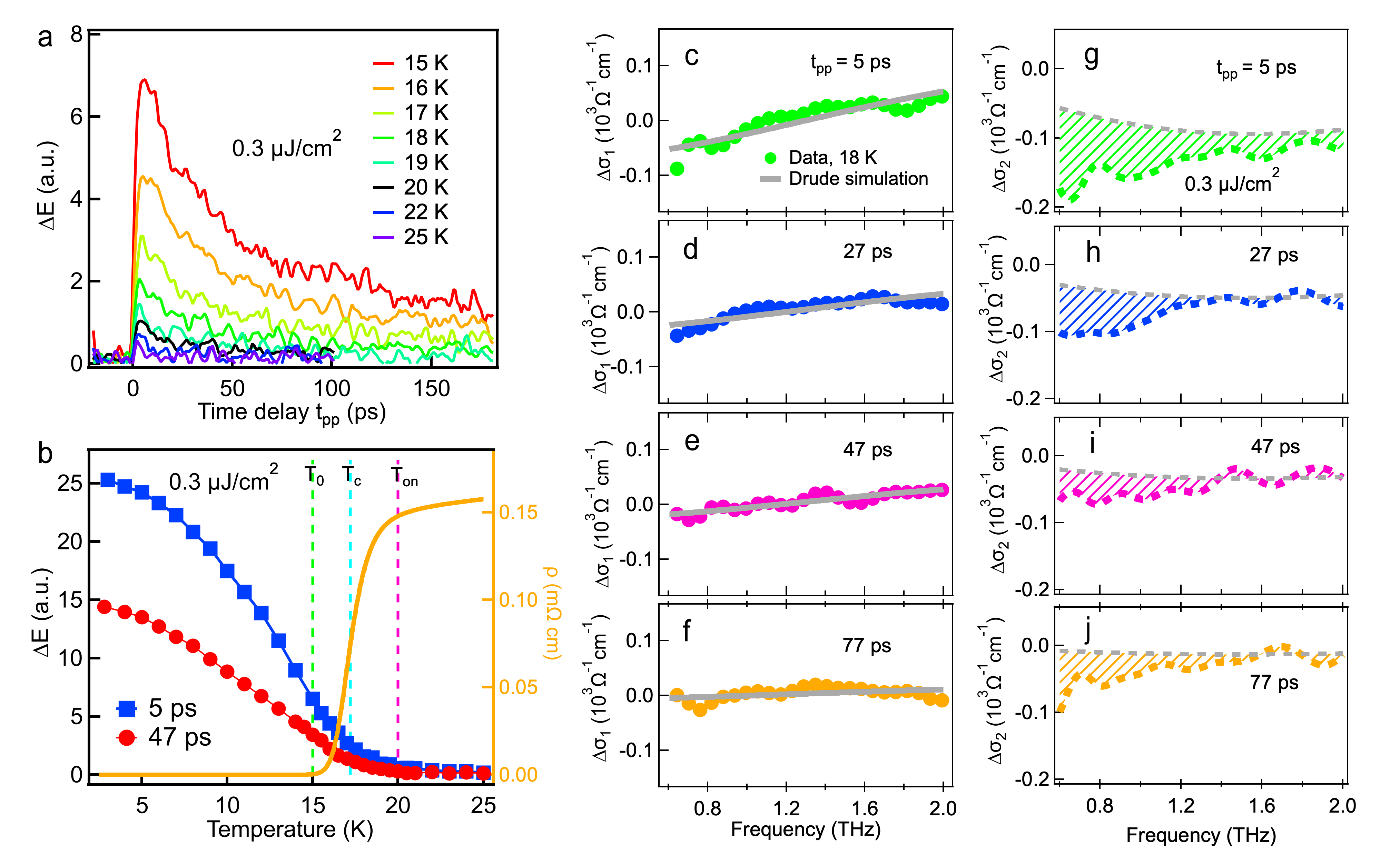}
\caption{ {\bf Superconducting fluctuation in the nickelate film.}  {\bf (a)} Pump-induced THz field change $\Delta E$ as a function of delay time $t_\mathrm{pp}$ above 15 K under a pump fluence of 0.3 $\mathrm{\mu J}$/cm$^2$. {\bf (b)} Pump-induced THz field change $\Delta E$ as a function of temperature at time delay $t_\mathrm{pp}$ = 5 and 47 ps under a pump fluence of 0.3 $\mathrm{\mu J}$/cm$^2$. DC resistivity $\rho(T)$ is also plotted together with $\Delta E(T)$. {\bf (c)} to {\bf (f)} The real part of ultrafast THz conductivity $\Delta\sigma_{1}(\omega)$ at $t_\mathrm{pp}$ = 5, 27, 47, 77 ps, respectively. {\bf (g)} to {\bf (j)} The imaginary part of ultrafast THz conductivity $\Delta\sigma_{2}(\omega)$ at $t_\mathrm{pp}$ = 5, 27, 47, 77 ps, respectively. The data were taken at 18 K under the pump fluence of 0.3 $\mathrm{\mu J}$/cm$^2$. The gray curves are pure Drude simulations to $\Delta\sigma_{1}$ and $\Delta\sigma_{2}$.}

\label{Fig4}
\end{figure*}

\noindent{\textbf{Superconducting gap 2$\Delta$ extracted by ultrafast superconducting kinetics}

After establishing the $d$-wave scenario, next we extract the $d$-wave gap amplitude in the nickelate film. In general, one way to determine $d$-wave gap amplitude is to map out the gap anisotropy in momentum space by using angle-resolved photoemission\cite{dwave_arpes_1993,ARPES_REVIEW_2003}. This kind of measurement is currently not feasible owing to the surface issue of nickelate films. An alternative approach which has been widely used to extract 2$\Delta$ in cuprates is the ultrafast optical pump-probe spectroscopy\cite{ultrafast_review}. It has been well-established that the ultrafast relaxation dynamics of photoexcited quasiparticles in cuprates is dominated by the recombination of antinodal excitations in the weak perturbation regime\cite{ultrafast_review,dwave_relaxation_theory_2004,dwave_relaxation_theory_2011}. After weak ultrafast excitations, multiple scattering and interaction between the superconducting condensate, pump-induced quasiparticles, and high-frequency phonons will lead to the establishment of a quasi-equilibrium. One could take advantage of the photoinduced signal amplitude, which is a measurement of the thermal quasiparticle density at the antinodal region, to extract the superconducting gap size at antinode\cite{dwave_gap_theory_2005}. The $d$-wave gap amplitude in cuprates extracted through this method is reasonably consistent with the values extracted by other techniques\cite{ultrafast_review,dwave_gap_theory_2005}.

In Fig. 3a and 3b, we present pump-probe delay time traces of $\Delta\sigma_1$ and $\Delta\sigma_2$ integrated over a 0.4 THz interval (centered at 0.95 THz) of the data in Fig. 2b and 2c respectively, and the delay-time scan of pump-induced THz electric field change $\Delta E$ at $t_\mathrm{gate}$ = 0 [inset, Fig. 3b]. The temporal evolution of $\Delta E(t_\mathrm{pp})$ accurately follows $\Delta\sigma_1(t_\mathrm{pp})$ and $\Delta\sigma_2(t_\mathrm{pp})$, demonstrating a direct correlation between the $\Delta E(t_\mathrm{pp})$ profile and the ultrafast dynamics of superconducting condensate. Thus, we fix time delay at $t_\mathrm{pp}$ $=$ 5 ps to maximize the pump-probe signal, and then measure $\Delta E$ as a function of temperature with high statistics. Relevant data under two weak pump fluence 0.04 and 0.07 $\mathrm{\mu J}$/cm$^2$ are shown in the inset of Fig. 3c. It is clearly visible that, as increasing temperature, $\Delta E$ gradually decreases and approaches zero at $\sim$ 20 K, behaving like an order parameter to characterize a superconducting phase transition. Based on the phenomenological Rothwarf and Taylor model, in the weak perturbation regime, i.e., the minimum depletion of superfluid density: $\Delta n/n_\mathrm{s}\ll$1, $\Delta E_{T\rightarrow0}$$/$$\Delta E$ $-1$ is proportional to the thermal quasiparticle density $n_T$\cite{dwave_gap_theory_2005} (see SI). We plot normalized $n_T$ as a function of temperature in Fig. 3c. One can see that $n_T$ under two weak pump fluences collapses to one curve, indicating our method allows a self-consistent extraction of $n_T$. The thermal quasiparticle density at temperature $T$ is determined by the equation $n_T$ $\simeq$ $N(0)$$\sqrt{2\pi\Delta k_\mathrm{B}T}$exp$(-\Delta/T)$\cite{ultrafast_review,dwave_gap_theory_2005}. Here $N(0)$ is the electronic density of states per unit cell and $\Delta$ is the superconducting gap in a \textit{d}-wave system. We use this equation to fit $n_T$ and results are shown in Fig. 3c. $n_T$ under two weak pump fluences are reasonably reproduced by the fitting. The gap amplitude 2$\Delta$ is extracted to be 1.2 THz (5 meV), resulting in a gap-to-$T_\mathrm{c}$ ratio 2$\Delta$$/k_\mathrm{B}T_\mathrm{c}$ $=$ 3.4. This ratio is similar to that observed in weak-coupling $s$-wave superconductors and in some $d$-wave electron-doped cuprates such as Pr$_{2-x}$Ce$_x$CuO$_{4-\delta}$\cite{ndoped_cuprates_ratio_2,n_doped_cuprates1,ndoped_cuprates_ratio_1}. However, it is slightly lower than the weakly coupled $d$-wave BCS value of about 4.3 that has been observed in heavily hole-doped cuprates like Bi$_2$Sr$_2$CaCu$_2$O$_{8+d}$\cite{Ratio_ovedoped_cuprates}. In this respect, the $d$-wave superconductivity in the nickelate film is also in the weak-coupling regime. It is worth noting that this gap size falls within the frequency range fully covered by our spectrometer. The absence of any hallmarks or onsets of quasiparticle excitations around 2$\Delta$ in equilibrium and ultrafast THz conductivity corroborates again a \textit{d}-wave scenario. We can estimate the ratio of  mean free path to the superconducting coherence length, $\l/\xi=\pi\tau\Delta/\hbar \sim$ 0.57, consistent with dirty-limit superconductivity.

\noindent{\textbf{Ultrafast THz probe of superconducting fluctuation near $T_\mathrm{c}$}}

At last, we discuss superconducting fluctuation in the nickelate film. To track possible fluctuating superconductivity with high precision, we used a slightly higher pump fluence $F$ = 0.3 $\mathrm{\mu J}$/cm$^2$ to quench superconductivity near $T_\mathrm{c}$. In Fig. 4a, we show pump-induced $\Delta E$ as a function of time delay above 15 K. Below 18 K, while the magnitude of $\Delta E(t_\mathrm{pp})$ is significantly decayed as temperature increases, the relatively long relaxation process indicates the pump-probe signal is mainly from superconductivity. Beyond 20 K, $\Delta E(t_\mathrm{pp})$ becomes negligible and its relaxation dynamics only limits to time zero, consistent with a weak normal metallic response. We further present temperature-dependent $\Delta E$ at two typical time delay $t_\mathrm{pp}$ $=$ 5 and 47 ps, and dc resistivity $\rho$ in Fig. 4b. Far below $T_\mathrm{c}$, $\Delta E$ shows a similar temperature dependence as the data in the inset of Fig. 3c. As temperature approaches $T_\mathrm{c}$, the decreasing trend of $\Delta E$ becomes weaker. Across $T_\mathrm{c}$, we still observe remarkable $\Delta E$. Above 20 K, $\Delta E$ of both time delays are flat and insensitive to temperature anymore. The dc resistivity of nickelate film defines three characteristic temperatures: ($i$) $T_0$ $\sim$ 15 K, the temperature of $\rho$ completely dropping to zero; ($ii$) $T_\mathrm{c}$ $\sim$ 17 K, the superconducting transition temperature; ($iii$) $T_\mathrm{on}$ $\sim$ 20 K, the temperature of $\rho$ deviating from a normal metallic behavior. A recent London penetration depth measurement using kHz inductance coils shows the superfluid density $n_\mathrm{s}$, a measurement of long-range coherence of Cooper pairs, completely vanishes near $T_0$\cite{nickelate_pen_depth_Huang_2022}, a temperature a few Kelvins below $T_\mathrm{c}$. This observation is consistent with our static THz measurement of $n_\mathrm{s}$ which drops to zero near $T_0$ as well (inset, Fig. 1b). Thus, $T_0$ could be roughly regarded as the onset temperature of long-range coherence of superconductivity. The low-frequency measurements such as KHz penetration depth measurement could determine the onset temperature of macroscopic coherence, whereas they cannot catch the temporal superconducting correlation in a timescale of picosecond. THz probe has been found much more sensitive to such temporal superconducting correlation\cite{SCfluctuation_nature_1999,SCfluctuation_2015}. We attribute the appreciable $\Delta E$ in the temperature region of $T_0$ $<$ $T$ $<$ $T_\mathrm{on}$ to the short-range superconducting fluctuation. Nickelate film in this temperature region is a fluctuating superconductor.

To reveal superconducting fluctuation in the nickelate film in a more quantitative way, we extracted $\Delta\sigma_1(\omega)$ and $\Delta\sigma_2(\omega)$ at four time delays $t_\mathrm{pp}$ = 5, 27, 47, 77 ps at 18 K. Relevant data are shown in Fig. 4c to 4j. One can see at all time delays, $\Delta\sigma_1(\omega)$ is weak and changes sign from low to high frequency. In contrast, $\Delta\sigma_2(\omega)$ is appreciable and always negative. Owing to the absence of the long-range superconducting phase at 18 K, we are able to use the Drude model to reproduce $\Delta\sigma_1(\omega)$ (see Method). It is clearly visible that the Drude fit has well captured the main features of $\Delta\sigma_1(\omega)$ at all delays. In stark contrast, $\Delta\sigma_2(\omega)$ cannot be described by the Drude model well. We highlight the residual $\Delta\sigma_2(\omega)$ which has subtracted the Drude contributions by shadowed area. Based on our analysis, we believe the residual $\Delta\sigma_2(\omega)$ highlighted by the shadowed area is parts of fluctuating conductivity from the short-range superconducting correlation.

\noindent{\textbf{Discussion and outlook}}

It is worth discussing similarities and differences of superconductivity in nickelate and cuprates. First of all, the $d$-wave gap amplitude in current nickelate film is found to be $\sim$ 1.2 THz, highlighting the superconductivity falls in the dirty limit (2$\Delta$ $<$ $\gamma_\mathrm{D}$). Thus, the superconductivity in nickelates is alike the dirty-limit $d$-wave superconductivity in electron-doped cuprates\cite{ndoped_cupreate_review_2010,LCCO_THzfilm_2021}. Second, our ultrafast THz probe detects superconducting fluctuation in nickelate film, but the superconducting fluctuation only exists in the temperature region where $T_\mathrm{c}$ is spread. In contrast, in optimally hole-doped cuprates like Bi$_2$Sr$_2$CaCu$_2$O$_{8+d}$, superconducting fluctuation could extend to 1.4 $T_\mathrm{c}$, indicating that the local superconducting pairing field survives in the deep normal state\cite{SCfluctuation_2015,SCfluctuation_2021}. Therefore, the superconducting fluctuation in optimally doped nickelate is intrinsically weaker than optimally hole-doped cuprates, providing further evidence of their analogy to the electron-doped cuprates. Third, because of the $d$-wave gap symmetry, a significant amount of uncondensed quasiparticles in a Drude-like peak has been both observed in nickelates and cuprates far below $T_\mathrm{c}$\cite{LSCO_infraredcrystal_2005,LSCO_THzfilm_2019,LCCO_THzfilm_2021}. The origin of these uncondensed quasiparticles, which has been intensively discussed in cuprates, is still highly controversial. One proposal to account these uncondensed quasiparticles adopts a dirty $d$-wave scenario in the framework of BCS theory\cite{curpate_uncondensed_theory_2018}. With considering the impurity scattering in a mixture of Born and unitary limit, the density of states near nodal region is significantly enhanced. Another possible mechanism considers the self-organized granularity in the disordered cuprates\cite{urpate_uncondensed_theory_2021}. As the coherence lengths of Cooper pairs and disorder potential become comparable, a simulation based on $t$-$J$ model predicts the superconducting pair-field amplitude in cuprates could be spatially heterogeneous. Both theories could reproduce the remarkable residual Drude component on conductivity spectra far below $T_\mathrm{c}$. Based on our observations, nickelate superconductor is providing a new platform to reexamine this puzzle originating from cuprates. Last but not the least, unlike the single-band cuprates, the nickelate superconductor is a multi-band system. Although extra three-dimensional electron pockets exist, neither equilibrium nor ultrafast THz conductivity reveal an $s$-wave component. These observations suggest that the $d$-wave gap symmetry is likely dominant on all Fermi surfaces. The superconducting gap extracted via optical methods represents an average pairing energy scale for the $d$-wave superconductivity in the nickelate film.

\bibliography{scibib}

\footnotesize

\bigskip

\section{Methods}

\textbf{Sample growth.}  $\sim$15 unit cells of perovskite Nd$_{0.85}$Sr$_{0.15}$NiO$_2$ thin films ($\sim$ 4.3 nm) were synthesized by pulsed-laser deposition with a KrF excimer laser ($\lambda$ = 248 nm) on 5 $\times$ 5 mm$^2$ LSAT (001) substrates, with the growth conditions specified in Ref. \cite{nickelate_LAST_2022}. The samples were then cut into two 2.5 $\times$ 5 mm$^2$ pieces and vacuum-sealed ($<$ 0.1 mTorr) in a Pyrex glass tube with $\sim$0.1 g of CaH$_2$ powder after loosely wrapping with aluminum foil. The glass tube was heated at 260 $^{\circ}$C for $\sim$2.5 hours, with temperature ramp rate of 10 $^{\circ}$C min$^{-1}$, to achieve topotactic transition to infinite-layer Nd$_{0.85}$Sr$_{0.15}$NiO$_2$\cite{nickelate_nature_2019,nickelate_LAST_2022}. The sample structural calibration information could be found in Ref. \cite{nickelate_LAST_2022}. These calibration data and dc resistivity measurements clearly demonstrate the nickelates films have excellent crystallinity and minimal defects.

\textbf{Static and ultrafast THz spectroscopy.}  A home-built optical pump THz probe spectrometer driven by a 1 kHz Ti:sapphire regenerative amplifier, which has 800 nm central wavelength and 40 fs pulse duration, was used for this work. Technical details can be found elsewhere\cite{THz_sc1,THztech}.  The laser is split into three beams.  One is used to optically pump the nickelate film. The other two beams are used to generate and detect phase-locked THz electric fields in time-domain via optical rectification and electro-optic sampling in a 1 mm thick ZnTe crystal, respectively. The transmitted THz probe containing spectral amplitude and phase information of the sample is directly measured in time domain using an optical gate pulse. The setup is enclosed in a dry air purge box. The static THz measurements were performed with the optical pump beam being fully blocked.

\textbf{Extraction of equilibrium terahertz conductivity.}  By mapping out THz pulses after transmitting through substrates and samples separately, and then performing Fourier transforms to get transmitted THz field of substrate ($E_\mathrm{sub}(\omega)$) and sample ($E_\mathrm{sam}(\omega)$) in frequency domain, we obtain complex transmission in the frequency domain [$T$($\omega$) = $E_\mathrm{sam}(\omega)$/$E_\mathrm{sub}(\omega)$]. The complex conductivity of thin films can be directly calculated in the thin-film limit with the expression: $T$($\omega$) = (1+$n$)/(1+$n$+$Z_0$$d$$\sigma$($\omega$))exp[i$\omega$/$c$($n$-1)$\Delta$$L$]. Here $T$($\omega$) is the complex transmission as referenced to a bare substrate. $\sigma (\omega)$ is the complex optical conductivity. $d$ is the thickness of the film, and $n$ is the index of refraction of the substrate. $\Delta L$ is the small thickness difference between sample and reference substrate. $Z_0$ is the vacuum impedance, which is approximately 377 $\Omega$.

\textbf{Extraction of ultrafast terahertz conductivity.}  To extract ultrafast terahertz conductivity $\Delta\sigma(\omega)$ of the nickelate film at pump-probe delay time $t_\mathrm{pp}$, we used the pump-probe scheme to measure pump-induced THz field change of sample $\Delta E_\mathrm{sam} (t)$, and then performed Fourier transforms to have $\Delta E_\mathrm{sam} (\omega)$. The complex transmission is obtained by $T_\mathrm{excited}(\omega)$ = [$\Delta E_\mathrm{sam} (\omega)$+$E_\mathrm{sam} (\omega)$]/$E_\mathrm{sub}(\omega)$. We then use the thin-film limit expression shown above to extract photoexcited transient THz conductivity $\sigma_\mathrm{excited} (\omega)$. The ultrafast terahertz conductivity could be extracted by $\Delta \sigma(\omega)$=$\sigma_\mathrm{excited} (\omega)$$-$$\sigma(\omega)$.

\textbf{Drude model fitting.}  To extract the normal state scattering rate, the terahertz conductivity data at 20 K was fit by a Drude model. We use one Drude oscillator to account for the normal Drude transport of free carriers. The expression is: 

\begin{equation}
\sigma(\omega)=\epsilon_0[\frac{\omega_p^2\tau_n}{1-i\omega\tau_n}-i(\epsilon_\infty-1)\omega]
\label{Drude1}
\end{equation}

\noindent Here, $1/2\pi\tau_n$ is the Drude scattering rate of the Drude term and $\omega_p$ is the Drude plasma frequency. The background polarizability $\epsilon_\infty$ originates from absorptions above the measured spectral range including phonons and interband absorptions.

\textbf{Two-fluid model fitting to equilibrium terahertz conductivity.}  In the superconducting state, we observed a Drude component from a large amount of uncondensed quasiparticles. To extract the superfluid density, we used a two-fluid model to reproduce the complex terahertz conductivity below $T_c$. The expression is: 

\begin{equation}
\sigma(\omega)=\epsilon_0[\frac{\omega_p^2\tau_n}{1-i\omega\tau_n}+\omega_s^2(\pi\delta(\omega)+\frac{i}{\omega})-i(\epsilon_\infty-1)\omega]
\label{Drude2}
\end{equation}

\noindent Here, $1/2\pi\tau_n$ is the Drude scattering rate of the Drude term and $\omega_p$ is the Drude plasma frequency. $\omega_s$ is the plasma frequency of superfluid condensate. The first term in the square brackets represents the Drude response of uncondensed quasiparticles. The second term represents the optical response of superfluid, i.e., a delta function in the real part of terahertz conductivity and a $1/\omega$ divergence in the imaginary part of terahertz conductivity. The background polarizability $\epsilon_\infty$ originates from absorptions above the measured spectral range including phonons and interband absorptions.  

\textbf{Two-fluid model fitting to ultrafast terahertz conductivity.}  In main text we showed optical pump-induced ultrafast terahertz conductivity $\Delta\sigma(\omega)=\Delta\sigma_1+i\Delta\sigma_2$ at 3 K. Our main observation is that upon a weak optical excitation, a small fraction of Cooper pairs are broken, resulting in a drop of $1/\omega$ term in imaginary terahertz conductivity and an increase of Drude response in real terahertz conductivity. To study the spectral weight transfer between superfluid density and photoexcited quasiparticles during this process, we used a  two-fluid model to fit the complex ultrafast terahertz conductivity. The expression is: 

\begin{equation}
\Delta\sigma(\omega)=\epsilon_0[\frac{\Delta\omega_p^2\tau_n}{1-i\omega\tau_n}-\Delta\omega_s^2(\pi\delta(\omega)+\frac{i}{\omega})]
\label{Drude3}
\end{equation}

\noindent Here, $1/2\pi\tau_n$ is the Drude scattering rate of the Drude term and $\Delta\omega_p$ is the Drude plasma frequency involved with the pair breaking process. $\Delta\omega_s$ is the plasma frequency of superfluid condensate pair breaking process. The first term in the square brackets represents the Drude response of broken Cooper pairs. The second term represents terahertz response of pump-depleted superfluid condensate.

\textbf{Drude model simulation to ultrafast terahertz conductivity at 18 K.}  In main text we showed that optical pump-induced ultrafast terahertz conductivity $\Delta\sigma(\omega)=\Delta\sigma_1+i\Delta\sigma_2$ at 18 K to reveal parts of superconducting fluctuating conductivity. As we showed in main text, at 18 K, the long-range coherence of superconductivity has already fully decayed so that we could use the Drude model to simulate the real part of terahertz conductivity. In main text Fig. 4c to 4f, the real part of ultrafast terahertz conductivity changes sign from low to high frequency. Because most superfluid condensate has already been thermally depleted, such a feature on real part of terahertz conductivity should come from the slight changes of Drude plasma frequency and scattering rate. To capture these features, first of all, we use a Drude fit to extract the Drude plasma frequency and scattering rate of equilibrium terahertz conductivity at 18 K. Then we use below formula to fit real part of ultrafast terahertz conductivity at different delays.

\begin{equation}
\sigma(\omega)=\epsilon_0[\frac{\omega_p^2\tau_n}{1-i\omega\tau_n}-\frac{\omega_{p,18\mathrm{K}}^2\tau_{n,18\mathrm{K}}}{1-i\omega\tau_{n,18\mathrm{K}}}]
\label{Drude4}
\end{equation}

\noindent Here, $1/2\pi\tau_n$ and $\omega_p$ are the plasma frequency and Drude scattering rate of the Drude term after the optical pumping respectively. $1/2\pi\tau_{n,18\mathrm{K}}$ and $\omega_{p,18\mathrm{K}}$ are the plasma frequency and Drude scattering rate of the Drude term without optical pumping at 18 K.

\textbf{Data availability.} All relevant data are available on reasonable request from J.W.

\section*{Acknowledgements}
This work was supported by the U.S. Department of Energy, Office of Basic Energy Science, Division of Materials Sciences and Engineering (Ames National Laboratory is operated for the U.S. Department of Energy by Iowa State University under Contract No. DE-AC02-07CH11358). B.C. was supported by the Laboratory Directed Research and Development project, Ames National Laboratory. Work at SIMES (K.L., Y.H.L., B.Y.W., Z.X.S., and H.Y.H.,) was supported by the U.S. Department of Energy, Office of Science, Basic Energy Sciences, Materials Sciences and Engineering Division under Contract No. DE-AC02-76SF00515 and the Gordon and Betty Moore Foundation’s Emergent Phenomena in Quantum Systems Initiative (Grant No. GBMF9072, synthesis equipment). Work at the University of Alabama, Birmingham (I.E.P.), was supported by the US DOE under contract no. DE-SC0019137.

\bigskip

\section*{Author Contributions}
B.C., Z.X.S., H.Y.H., and J.W. initiated the project; B.C. and D.C. performed the measurements with the help of L.L.; K.L., Y.H.L., B.Y.W., and H.Y.H. developed samples and performed transport characterizations; B.C. and J.W. analyzed the spectroscopy data with the help of L.L., Z.Y.C., M.M., and I.E.P.; The manuscript was written by B.C. and J.W. with input from all authors; J.W. supervised the project.

\section{Additional information:} 
 
\textbf{Supplementary Information} accompanies this paper at

\bigskip
 
\textbf{Competing financial interests:} The authors declare no competing financial interests.
 
\bigskip
 
\textbf{Reprints and permission} information is available online at 
\bigskip

\normalsize

\end{document}